\documentclass[sigconf]{acmart}
\AtBeginDocument{%
  }

\setcopyright{acmlicensed}
\copyrightyear{2018}
\acmYear{2018}
\acmDOI{XXXXXXX.XXXXXXX}
\acmConference[Conference acronym 'XX]{Make sure to enter the correct
  conference title from your rights confirmation email}{June 03--05,
  2018}{Woodstock, NY}
\acmISBN{978-1-4503-XXXX-X/2018/06}

\usepackage{multirow}
\usepackage{multicol}
\usepackage{amsmath}
\usepackage{xcolor}
\usepackage{graphicx}
\usepackage{makecell}
\usepackage{subfigure}
\usepackage{enumitem}
\newcommand{\modelname}{MISS }

\begin{document}

\title{MISS: Multi-Modal Tree Indexing and Searching with Lifelong Sequential Behavior for Retrieval Recommendation}

\author{Chengcheng Guo}
\authornote{Equal contribution.}
\affiliation{%
\country{Kuaishou Inc., Beijing, China}
}
\email{guochengcheng03@kuaishou.com}

\author{Junda She}
\authornotemark[1]
\affiliation{%
\country{Kuaishou Inc., Beijing, China}
}
\email{shejunda@kuaishou.com}

\author{Kuo Cai}
\affiliation{%
\country{Kuaishou Inc., Beijing, China}
}
\email{caikuo@kuaishou.com}

\author{Shiyao Wang}
\affiliation{%
\country{Kuaishou Inc., Beijing, China}
}
\email{wangshiyao08@kuaishou.com}

\author{Qigen Hu}
\affiliation{%
  \country{Kuaishou Inc., Beijing, China}
}
\email{huqigen03@kuaishou.com}

\author{Qiang Luo$^{\dag}$}
\affiliation{%
\country{Kuaishou Inc., Beijing, China}
}
\email{luoqiang@kuaishou.com}

\author{Kun Gai}
\affiliation{%
\country{Unaffiliated, Beijing, China}
}
\email{gai.kun@qq.com}

\author{Guorui Zhou}
\authornote{Corresponding author.}
\affiliation{%
\country{Kuaishou Inc., Beijing, China}
}
\email{zhouguorui@kuaishou.com}

\renewcommand{\shortauthors}{ChengCheng Guo, Junda She et al.}

\begin{abstract}
Large-scale industrial recommendation systems typically employ a two-stage paradigm of retrieval and ranking to handle huge amounts of information.
Recent research focuses on improving the performance of retrieval model.
A promising way is to introduce extensive information about users and items.
On one hand, lifelong sequential behavior is valuable.
Existing lifelong behavior modeling methods in ranking stage focus on the interaction of lifelong behavior and candidate items from retrieval stage.
In retrieval stage, it is difficult to utilize lifelong behavior because of a large corpus of candidate items.
On the other hand, existing retrieval methods mostly relay on interaction information, potentially disregarding valuable multi-modal information.
To solve these problems, we represent the pioneering exploration of leveraging multi-modal information and lifelong sequence model within the advanced tree-based retrieval model. 
We propose \textbf{M}ulti-modal \textbf{I}ndexing and \textbf{S}earching with lifelong \textbf{S}equence (\textbf{MISS}),
which contains a multi-modal index tree and a multi-modal lifelong sequence modeling module.
Specifically, for better index structure, we propose multi-modal index tree, which is built using the multi-modal embedding to precisely represent item similarity.
To precisely capture diverse user interests 
in user lifelong sequence, we propose collaborative general search unit (Co-GSU) and multi-modal general search unit (MM-GSU) for multi-perspective interests searching. 
Online experiments have demonstrated the effectiveness of the proposed method. 
\end{abstract}




\maketitle

\section{Introduction}
To balance efficiency and effectiveness, a cascade framework, which consists of retrieval (also called matching, recall, \textit{etc}.) and ranking stages, has been widely adopted in modern recommendation systems \cite{gao2019learning}. 
Within this framework, retrieval stage is crucial in distinguishing candidates from the vast corpus but given the least time.
DSSM\cite{huang2013dssm} is a classic dual-tower architecture, which employs the Approximate Nearest Neighbor (ANN) method for efficient item retrieval.
Within the dual-tower paradigm, user representation and item representation can only interact at a later stage.
For early interaction, a line of work seeks to design index structure for candidate retrieval. 
Tree-based Deep Models (TDM\cite{zhu2018tdm}, JDM\cite{zhu2019jdm}, BSAT\cite{zhuo2020bsat}) propose a tree-based index to hierarchically retrieve candidates.
NANN\cite{chen2022nann} resorts to HNSW\cite{malkov2018hnsw}, a graph-based index.
Deep Retrieval (DR)\cite{gao2021DeepRetrieval} defines items into "paths".
Recently, Streaming VQ\cite{bin2025streamingVQ} utilizes vector quantization as a new index structure.
With the index structure, complicated models originally designed for the ranking stage can be used in the retrieval stage for representations interaction between users and items.

Among complicated models in the ranking stage, lifelong sequential behavior modeling methods (e.g. SIM) are effective.
The interactions of lifelong sequence and candidate items from retrieval stage using complicated models are effective to model user interests.
SIM\cite{pi2020sim} is a search-based method, which introduces a General Search Unit (GSU) for behavior searching and an Exact Search Unit (ESU) for precise relationship modeling.
TWIN\cite{chang2023twin} extend the target attention structure to GSU, thus addressing the inconsistency between GSU and ESU.
In retrieval stage, a large corpus of candidate items makes it difficult to utilize the lifelong sequence.

The above methods depend on behavior information, potentially disregarding valuable multi-modal information. 
With the significant evolution of multi-modal large models, the potential of multi-modal information in the recommendation system has received great attention\cite{liu2024multimodal, liu2024multimodal-survey-kdd}. 
AlignRec\cite{liu2024alignrec} trains multi-modal embeddings using visual-text alignment task and collaborative filtering task.
\citet{sheng2024taobao} proposes semantic-aware contrastive learning to pre-train multi-modal embeddings.
QARM\cite{luo2024qarm} aligns multi-modal embeddings with item-item relationships in pre-training and leverages quantitative code of multi-modal embeddings for recommendation. 
Above all, existing multi-modal work mostly concentrates on the learning of multi-modal embeddings.

To tackle these problems, we propose \textbf{M}ulti-modal \textbf{I}ndexing and \textbf{S}earching with lifelong \textbf{S}equence (\textbf{MISS}),
a pioneer model that introduces lifelong sequence model and multi-modal information into tree-based model. 
To be specific, \modelname contains two components: a multi-modal index tree and a multi-modal lifelong sequential behavior modeling module.
Since this paper does not focus on the training of multi-modal embeddings, we directly use item alignment\cite{luo2024qarm} to train multi-modal embeddings, which contain content and interaction information.
As for the index tree, it is essential to organize similar items in the same subtree\cite{zhu2018tdm}.
Thus, we construct the tree using multi-modal embeddings to precisely reflect item similarity.
As for modeling lifelong behavior, it is crucial to extract diverse interests in user lifelong sequence.
To achieve this goal, we propose a collaborative general search unit (Co-GSU) and a multi-modal general search unit (MM-GSU), which search user interests with collaborative and multi-modal information.
In fact, \modelname has been serving in Kuaishou's recommendation system, leading to remarkable user engagement gain.
Our contributions can be summarized as follows:
\begin{itemize}[leftmargin=*]
    \item To the best of our knowledge, our work represents the pioneering exploration of leveraging multi-modal information within advanced retrieval models.
    \item We propose a multi-modal tree-based deep model, which contains a multi-modal index tree and a lifelong sequence learning modulde with Co-GSU and MM-GSU.
    \item We conduct extensive experiments to demonstrate the effectiveness of our method. Our method is also validated through online A/B test, showing promising results.
\end{itemize}

\section{Related Works}

\subsection{Indexing}
In the retrieval stage, indexing schemes are essential for effectively organizing and retrieving large-scale items\cite{huang2024survey-retrieval}.
DSSM\cite{huang2013dssm} is so-called 'two tower model', which searches by Approximate Nearest Neighbor (ANN) method.
Tree-based Deep Models (TDM\cite{zhu2018tdm}, JDM\cite{zhu2019jdm}, BSAT\cite{zhuo2020bsat}) propose tree structures to hierarchically search candidates from coarse to fine and make decisions for each user-node pair.
NANN\cite{chen2022nann} resorts to HNSW\cite{malkov2018hnsw} for candidate searching.
To avoid the Euclidean space assumption in the ANN algorithms, Deep Retrieval (DR)\cite{gao2021DeepRetrieval} defines items into 'paths' and uses beam search to shrink candidates layer by layer.
Recently, Streaming VQ\cite{bin2025streamingVQ} utilizes vector quantization as a novel index structure.

\subsection{Lifelong Sequential Behavior Modeling}
User Long Sequence Modeling is vital for capturing user interests \cite{pi2020sim,he2023survey-longseq}. MIMN\cite{pi2019mimn} integrates user behaviors learning with serving systems. SIM \cite{pi2020sim} introduces a General Search Unit (GSU) for behavior retrieval and an Exact Search Unit (ESU) for precise relationship modeling. EAT \cite{chen2021eta} uses locality-sensitive hashing for item embeddings and Hamming distance for retrieval. SDIM \cite{cao2022SDIM} samples items with the same hash signature, aggregating these via ESU to derive user interests. TWIN \cite{chang2023twin} extends the target attention structure to GSU and synchronizes embeddings and attention parameters between ESU and GSU.

\subsection{Multi-Modal Recommendation}
DVBPR\cite{kang2017DVBPR} jointly trains a CNN visual encoder with the Matrix Factorization task.
BM3\cite{zhou2023BM3} leverages self-supervised learning to align both the inter-modality and intra-modality representations within the collaborative filtering task.
AlignRec\cite{liu2024alignrec} pre-trains the visual-text alignment task and then aligns the multi-modal representations and the ID representations supervised by collaborative filtering task.
\citet{sheng2024taobao} proposes semantic-aware contrastive learning in the pre-training stage and extracts features using SimTier and MAKE based on fixed multi-modal representations.
QARM\cite{luo2024qarm} aligns multi-modal representation with down-stream business-specific item-item relationships in pre-training and leverages quantitative code mechanisms for end-to-end training in recommendation models.

As mentioned above, existing work mostly focuses on the pre-training stage to learn better multi-modal representations.
However, how to effectively utilize multi-modal information in the retrieval recommendation has rarely been explored.

\begin{figure*}[htbp]
    \centering
    \includegraphics[width=\textwidth]{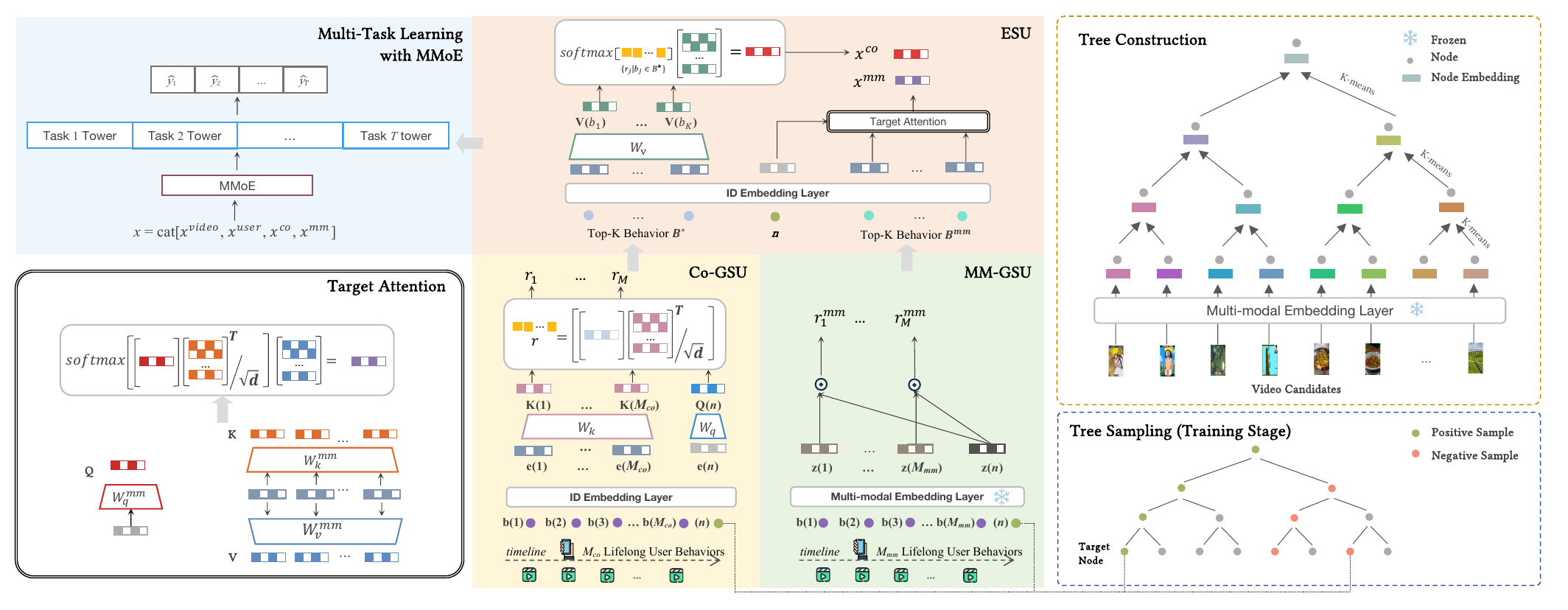}\vspace{-0.3cm} 
    \caption{The overall framework of MISS. 
    a) Tree Construction: the multi-modal index tree is first built based on multi-modal embeddings using k-means. b) 
    Tree Learning: given a positive or negative sample and lifelong user behaviors, the Co-GSU and MM-GSU would calculate the relevant scores $r_i$ and $r^{mm}_i$, respectively. Top-K behaviors are selected as the inputs of ESU. In ESU, target attention is performed to get two representations $x^{co}$ and $x^{mm}$. Finally, multiply representations are sent into MMoE module for 'pxtr' prediction.}
    \label{fig:model} 
\end{figure*}

\section{Preliminary}
\textbf{Problem Definition.} Let $\mathcal{U}$ and $\mathcal{V}$ denote the sets of users and videos, respectively, with the size of the video set given by $|\mathcal{V}|$. During the retrieval phase, for any given user $u\in \mathcal{U}$, we aim to extract a small subset of videos $\mathcal{V}_u \subset \mathcal{V}$ that align with user interests, satisfying $|\mathcal{V}_u| \ll V$.
In our model $\mathcal{M}$, we employ a multi-objective learning approach to optimize it. For videos $v$ watched by user $u$, user feedback—likes, comments, playing completions—is captured in a binary vector $\{y_1, \ldots, y_T\} \in \mathbb{R}^T$, where each $y_i \in \{0, 1\}^V$. The model learns from these feedbacks across $T$ different objectives. Each video is also represented by a multi-modal feature vector $m$. Our model is structured as a tree, comprising an index tree $\mathcal{T}$ and a node estimator $\mathcal{F}_\theta$. Following \cite{deng2025csmf}, we denote the training dataset for task $t$ as $\mathcal{D}_{tr}^t$ and the testing dataset as $\mathcal{D}_{ts}$.

\textbf{Multi-Modal Embedding.} 
To introduce behavior information into content-based representations, we follow the item alignment mechanism\cite{luo2024qarm} to supervise the fine-tuning of content-based embedding with knowledge from retrieval models.
Specifically, we select item pairs $(i,j)$ with high similarity from our item2item retrieval model as the data source $\mathcal{D}$, then train a multi-modal representation model (MRM) with pure content-based inputs (images and text) and interaction-based supervision signals.
For a random data batch in $\mathcal{D}$, we select in-batch negative samples $\mathcal{J}$ for training. 
\begin{equation}
    c_i = \mathbf{MRM}(c^{text}_i,c^{image}_i),\ 
    c_j = \mathbf{MRM}(c^{text}_j,c^{image}_j),
\end{equation}
\begin{equation}
    \mathcal{L}_{IA} = \mathbf{InfoNCE}(c_i,c_j,\mathcal{J}),    
\end{equation}
where $c_i$ and $c_j$ are the generated multi-modal embeddings, and $\mathbf{InfoNCE}$ pulls positive samples closer while pushing negatives away.
As a result, the so-called multi-modal embedding contains both content modality and interaction modality.

\textbf{Index Tree.} Let $\mathcal{T}$ be a binary tree of height $H$, with the root at level 0 and leaves at level $H$. We denote nodes at level $h$ as $\mathcal{N}_h$ and all nodes as $\mathcal{N} = \bigcup_{h=0}^H \mathcal{N}_h$. Each leaf node corresponds uniquely to a video, establishing a bijective mapping $\pi:\mathcal{N}_H \to \mathcal{V}$.
For each node $n \in \mathcal{N}$, we denote leaf nodes in its subtree by $\mathbf{Leaf}(n)$ and its ancestor at level $h$ as $\mathbf{Anc}(n;h)$.

\section{Methods}
\subsection{Multi-modal Index Tree}

In this section, we build the multi-modal index tree $\mathcal{T}$ using our multi-modal embedding because of its outstanding properties.
As mentioned in TDM\cite{zhu2018tdm}, the tree should represent hierarchical information of user interests, and it is natural to build the tree in a way that similar items are organized in close position. 
Recall that the multi-modal embeddings are trained to pull similar items closer using content representation.
Thus, the multi-modal embeddings are equipped to represent multidimensional similarity, both in terms of content and interactions.
Experimental results show the effectiveness of our multi-modal index tree.

\textbf{Tree Construction.} With the multi-modal embedding, we construct a tree using the k-means algorithm. 
At each step, items are divided into two clusters according to their multi-modal embeddings. 
Subsequently, similar operations are conducted within each of the clusters.
The recursion stops when only one item is left and a binary tree could be constructed in such a top-down way.

Each non-leaf node is a cluster center and its embedding is the mean pooling of the multi-modal embedding of the leaf node in its subtree.
Formally, the multi-modal embedding for each node is as follow:
\begin{equation}
    z_i =
    \begin{cases}
        c_{\pi(i)}, & \text{if } i \in \mathcal{N}_H , \\
        \mathbf{MeanPooling}(\{c_{\pi(j)}|j \in \mathbf{Leaf}(i)\}), & \text{else}.
    \end{cases}
\end{equation}
where $\pi(i)$ is the corresponding video of node $i$, $\mathcal{N}_H$ is the set of leaf node, $\mathbf{MeanPooling}$ is the mean pooling operation, $\mathbf{Leaf}(i)$ is the leaf node set of the subtree of node $i$.

\textbf{Learning of Tree Model.} 
To make the tree a max-heap\cite{zhu2018tdm}, we first give each node an ID embedding and train the model $\mathcal{F}_\theta:\mathcal{V}\times\mathcal{N}\to\mathbb{R}$ with node-wise task.
For any instance $(u,\mathbf{y})$, a pseudo label $\tilde{y}_n\in\{0,1\}$ is defined for each node $n \in \mathcal{N}$ to represent the existence of relevant targets on the subtree of $n$, i.e.,
\begin{equation}
    \tilde{y}_n = \mathbb{I}(\{\sum_{i\in \mathbf{Leaf}(n)} y_{\pi(i)} \geq 1 \}),
\end{equation}

With the pseudo labels, we can train the node estimator $\mathcal{F}_\theta$ level by level and negative sampling would be used for each level.
For level $h$ of the tree, the positive target set is defined as $\mathcal{S}^{+}_h(y)=\{n: \tilde{y}_n=1,n \in \mathcal{N}_h\}$.
$\mathcal{S}^-_h(y)$ contains several negative samples from $\{n: \tilde{y}_n=0,n \in \mathcal{N}_h\}$. 
The sample set for level $h$ is defined as $\mathcal{S}_h(y)=\mathcal{S}^{+}_h(y)\bigcup\mathcal{S}^{-}_h(y)$.
Formally, the training loss is as follow:
\begin{align}
    \mathcal{L} &=\sum_{t=1}^T \sum_{(u,y) \in \mathcal{D}_{tr}^t} \sum_{h=1}^{H} \sum_{n\in \mathcal{S}_h(y)} \mathcal{L}_{BCE}(\tilde{y}_n,\hat{y}_n^t), \label{Eq. final loss} \\
    \mathcal{L}_{BCE}(\tilde{y}_n,\hat{y}_n^t) &=-\tilde{y}_nlogy-(1-\tilde{y}_n)log(1-\hat{y}_n^t), \label{Eq. bce loss} \\
    \hat{y}_n^t &= \mathcal{F}_{\theta}(u,n;t),
\end{align}
where $\mathcal{F}_\theta$ is the node estimator. $\mathcal{F}_\theta$ contains a multi-modal lifelong sequence learning module and a multi-task learning module, which are introduced in Section \ref{longSeq} and Section \ref{MMOE} respectively.

Unlike TDM\cite{zhu2018tdm}, we would not build a new index tree using the trained ID embeddings, because the ID embeddings are trained just with interaction data and the absence of content information would lead to suboptimal results.
The experimental results have also demonstrated the effectiveness of our multi-modal index tree.

\textbf{Inference of Tree Model.}
When inferring, we use the beam search to sample multiply nodes.
For any testing instance $(u,\mathbf{y})\in \mathcal{D}_{ts}$, suppose $\mathcal{B}_h^{(K)}(u)$ denotes the node set retrieved at level $h$ through beam search and $K=|\mathcal{B}_h^{(K)}(u)|$ denotes the beam size, the beam search process is defined as:
\begin{align}
    \mathcal{B}_h^{(K)}(u) \in \underset{n\in\tilde{\mathcal{B}}_h(u)}{\text{argTopK }} p_\mathcal{F}(\tilde{y}_n=1|u),
    \label{beam}
\end{align}
where $\tilde{\mathcal{B}}_h(u)=\bigcup_{n\in\mathcal{B}_{h-1}^{(K)}(u)}\mathcal{C}(n)$, $p_\mathcal{F}(\tilde{y}_n=1|u)$ is the possibility from node estimator $\mathcal{F}$.
By applying Eq.\ref{beam} recursively until $h=H$, beam search retrieves the set containing $K$ leaf nodes, denoted by $\mathcal{B}_H^{(K)}(u)$.
Let $m \leq K$ be the retrieval number, the retrieval target set is defined as follow:
\begin{align}
    \hat{\mathcal{V}}_u &= \{\pi(b):n\in \mathcal{B}_H^{(m)}(u)\}, \\
    \mathcal{B}_H^{(m)}(u) &= \underset{n\in\mathcal{B}_H^{(K)}(u)}{\text{argTopm }} p_\mathcal{F}(\tilde{y}_n=1|u),
\end{align}


\subsection{Multi-Modal Searching with Lifelong Sequence}
\label{longSeq}
In this section, we introduce user lifelong behavior sequence in the retrieval stage for better user modeling.
Advanced researches\cite{pi2019mimn,pi2020sim} show that considering long-term historical behavior sequences in user interest modeling can significantly improve prediction performance of XTR".
Although a longer user behavior sequence introduces useful information about user interest, it contains massive noise at the same time\cite{qin2020UBR4CTR,pi2020sim,he2023survey-longseq}.
To mitigate the influence of noise and precisely capture user interest, we introduce two kinds of general search unit (GSU): multi-modal general search unit (MM-GSU) and collaborative general search unit (Co-GSU).
MM-GSU and Co-GSU would retrieve relevant behavior based on multi-modal information and collaborative information, respectively. Then an exact search unit (ESU) models the precise relationship between the candidate TDM node and the retrieved behaviors.

\textbf{Co-GSU.} Given the list of user behavior $B=[b_1;b_2;...;b_{M_{co}}]$ and a candidate node $n$ from $\mathcal{T}$ (either virtual node or real item node), each $b_i$ and $n$ are first denoted as one-hot vectors and then embedded into low-dimensional vectors $E=[e_1;e_2;...;e_{M_{co}}]$ and $e_n$ through the ID embedding layer.
After that, we take advantage of target attention\cite{zhou2018din} to calculate the relevant score:
\begin{align}
    r_i = (W_q e_n)^T \cdot (W_k e_i)/\sqrt{d},
\end{align}
where $W_q$ and $W_k$ are the parameters of query matrix and key matrix, $e_i$ and $e_n$ are denoted as the ID embeddings of the $i$-th behavior $b_i$ and the candidate node $n$ respectively, d is the dimension of ID embeddings.
After that, the Top-K relevant behaviors are selected as a \textbf{Co}ollaborative \textbf{S}ub user \textbf{B}ehavior \textbf{S}equence (Co-SBS) $B^{\star}$. 

\textbf{MM-GSU.} Given the list of user behavior $B=[b_1;b_2;...;b_{M_{mm}}]$ and a candidate node $n$ from $\mathcal{T}$ (either virtual node or real item node), the lookup operation converts each $b_i$ and $n$ into multi-modal embeddings $Z=[z_1;z_2;...;z_{M_{mm}}]$ and $z_n$. 
It is worth noticing that the multi-modal embeddings are well pre-trained and frozen in MM-GSU.
We then use the multi-modal embeddings to calculate the relevant score $r_i$ for each behavior $b_i$. 
Unlike the Co-GSU, we directly utilize the multi-modal embedding of candidate node as query and multi-modal embedding of each behavior $b_i$ as key.
The reason is that the multi-modal embeddings are trained to pull similar items closer and have formed a Euclidean space which suits calculating relevant scores.

Formally, the relevant score is calculated as $r_i^{mm} = (z_n)^T \cdot z_i$,
where $z_i$ and $z_n$ are denoted as the multi-modal embeddings of the $i$-th behavior $b_i$ and the candidate node $n$ respectively. 
After that, the Top-K relevant behaviors are selected as a \textbf{M}ulti-\textbf{M}odal \textbf{S}ub user \textbf{B}ehavior \textbf{S}equence (MM-SBS) $B^{mm}=[b_1,b_2,...,b_K]$. 


\textbf{ESU.} With the MM-SBS and Co-SBS, the exact search unit (ESU) applies target attention between the candidate node and the two subsequences, respectively.
It is worth noticing that the attention module of MM-SBS, Co-SBS and Co-GSU shares parameters.

Recalling that we have calculated the relevant scores in Co-GSU, we can reuse the scores to reduce the computation. We directly compute \textit{\textbf{softmax}} on the scores of Co-SBS to get the attention scores. 
\begin{equation}
    a_i = softmax(\{r_j|b_j\in B^{\star}\})_i,\ 
    x^{co} = \sum_{b_i \in B^{\star}} a_i W_v e_i.
\end{equation}
where $W_v$ are the parameters of projection matrices.

As for MM-SBS $B^{mm}=[b_1,b_2,...,b_K]$, each $b_i$ and $n$ are first denoted as one-hot vectors and then embedded into low-dimensional vectors $E^{'}=[e_1;e_2;...;e_K]$ and $e_n$ through the ID embedding layer. 
\begin{align}
    a_i &= softmax(\{(W_q^{mm} e_n)^T \cdot (W_k^{mm} e_j)/\sqrt{d}|e_j\in E^{'}\})_i, \\
    x^{mm} &= \sum_{b_i \in B^{mm}} a_i W_v^{mm} e_i.
\end{align}
where $W_q^{mm}$, $W_k^{mm}$ and $W_v^{mm}$are the parameters of projection matrices, which are not shared  parameters with Co-SBS.

\subsection{Multi-Task Learning with MMoE}
\label{MMOE}
The collaborative behaviors representation and multi-modal behaviors representation are subsequently concatenated with user feature and video feature of target item to construct the input of Multi-gate Mixture-of-Experts module (MMoE)\cite{ma2018mmoe}.
In MMoE module, we incorporate $L$ experts and introduce a separate gating network $g^t$ for each objective $t$. 
Then individual prediction $\hat{y}^t$ for each objective follows the corresponding objective-specific tower $h^k$.
Formally, the output of objective $t$ is 
\begin{align}
    \hat{y}^t_n &= h^t(f^t(x)) ,\ 
    f^t(x) = \sum_i^L g^t(x)_i f_i(x), \\
    x &= concat(x^{video}, x^{user}, x^{co}, x^{mm}).
\end{align}
where $f_i$ is the expert network, $g^t(x)$ is the gated network, $g^t(x)_i$ is the $i$-th logit of $g^t(x)$, $h^t$ is the tower network of objective $t$.

With the predicted possibility $\hat{y}^t_n$, the final loss can be calculated with Eq. \ref{Eq. final loss}-\ref{Eq. bce loss}.
During inference, we would use the mean of the prediction scores of all objectives as the final prediction score.

\begin{figure}[htbp]
    \centering
    \includegraphics[width=0.45\textwidth]{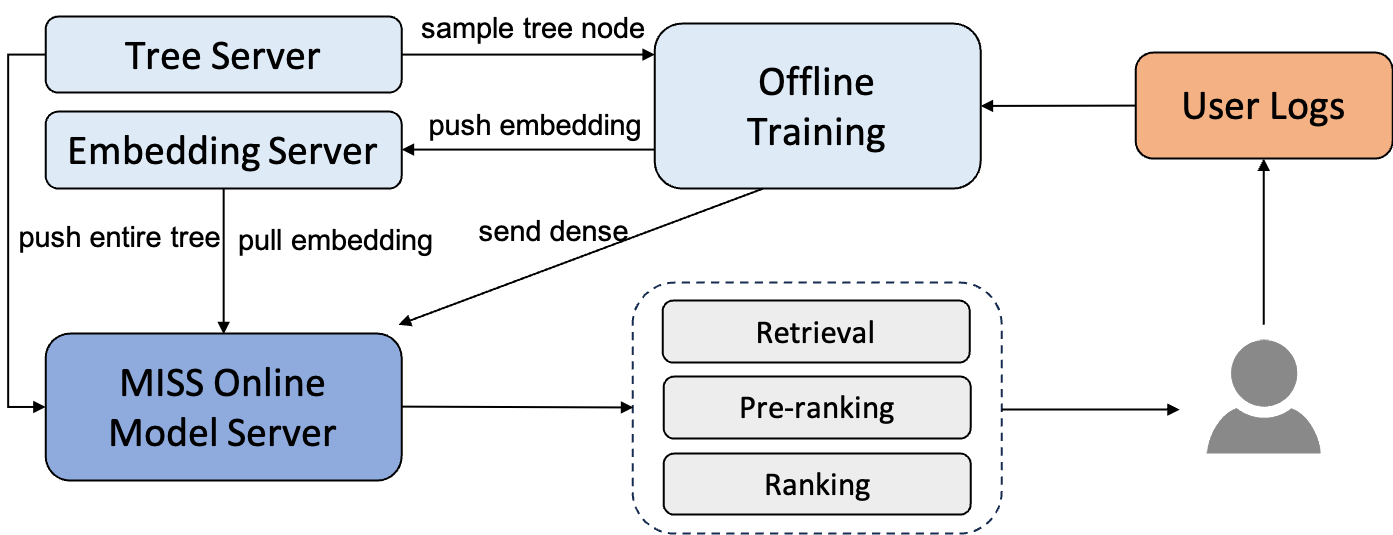} 
    \caption{System deployment of the proposed method. 
    }
    \label{fig:sys}\vspace{-0.2cm}
\end{figure}

\section{System Deployment}
We deploy MISS on the retrieval stage of Kuaishou, serving 400 million daily active users.In this section, we introduce the deployment of MISS at Kuaishou. Details of our system architecture is shown in Figure 2.

\textbf{Offline Training.} Our proposed MISS is trained on a large-scale distributed learning system of Kuaishou. Every day, Hundreds of millions of users watch short videos at Kuaishou APP, generating tens of billions of training logs. Before training, Tree Server performs kmeans construction on the pre-trained video multimodal embedding. During the training phase, the trainer requests Tree Server to pull the tree node representation and update the sparse features to the embedding server.

\textbf{Online Serving.} MISS Online Model Server loads the neural network parameters and the entire tree structure sent by the Tree Server at the initialization stage, and performs BeamSearch to retrieve video results in the online stage. MISS takes effect in the retrieval stage of the Kuaishou recommendation system, and is filtered with other retrieval channels in the Pre-ranking and Ranking stages, and finally displayed to users.

\begin{table*}[htbp]
\centering
\caption{Final retrieval results using recall as metric. The best and second-best results highlighted in \textbf{bold} font and \underline{underlined}. 
}\vspace{-0.2cm}
\resizebox{0.9\linewidth}{!}{
\begin{tabular}{c|c|c|c|c|c|cc|c}
\hline
\multirow{2}{*}{metric} & \multicolumn{7}{c|}{method} & \multirow{2}{*}{improvement} \\ \cline{2-8} 
& {SASRec} & {NANN} & {Kuaiformer} & {TDM} & {TDM+MMoE} & {MISS (2k seq)} & {MISS (4k seq)}  \\ \hline
Recall@800  & 0.20±0.01305   & 0.23±0.01875 & 0.26±0.01531       & 0.22±0.02420 & \underline{ 0.29±0.01558} & 0.38±0.01698 & \textbf{0.40±0.01232} & 37.93\% \\ \hline
Recall@600  & 0.18±0.00585 & 0.20±0.01301 & 0.23±0.01111       & 0.19±0.02575 & \underline{ 0.26±0.01451} & 0.33±0.01373 & \textbf{0.34±0.01607} & 30.77\% \\ \hline
Recall@400 & 0.15±0.00516  & 0.15±0.01972 & \underline{ 0.19±0.02039} & 0.14±0.02796 & 0.18±0.02562       & 0.26±0.02220 & \textbf{0.28±0.01570} & 47.37\%\\ \hline
\end{tabular}
}\vspace{-0.3cm}
\label{table:main}
\end{table*}

\begin{table}[htbp]
\centering
\caption{Retrieval results of intermediate nodes selected during beam search. We analyze the levels 13, 16 and 19 using hierarchical recall metric. The best and second-best results in each column are highlighted in \textbf{bold} font and \underline{underlined}. 
}
\resizebox{0.99\linewidth}{!}{
\begin{tabular}{c|c|c|cc}
\hline
metric         & TDM          & TDM+MMoE           & MISS (2k seq)     & MISS (4k seq)              \\ \hline
H@13Recall@800 & 0.76±0.01936 & \underline{ 0.76±0.01504} & 0.83±0.01703 & \textbf{0.85±0.01366} \\ 
H@16Recall@800 & 0.38±0.02100 & \underline{ 0.42±0.01310} & 0.60±0.01223 & \textbf{0.64±0.01003} \\
H@19Recall@800 & 0.26±0.02132 & \underline{ 0.32±0.01507} & 0.46±0.01775 & \textbf{0.51±0.01426} \\ \hline
H@13Recall@600 & 0.66±0.01922 & \underline{ 0.70±0.01015} & 0.80±0.01526 & \textbf{0.81±0.00975} \\
H@16Recall@600 & 0.31±0.02091 & \underline{ 0.38±0.01701} & 0.56±0.01008 & \textbf{0.58±0.00961} \\
H@19Recall@600 & 0.21±0.01979 & \underline{ 0.28±0.01014} & 0.42±0.01439 & \textbf{0.46±0.01477} \\ \hline
H@13Recall@400 & 0.55±0.02721 & \underline{ 0.58±0.02353} & 0.70±0.01380 & \textbf{0.72±0.01091} \\
H@16Recall@400 & 0.24±0.02563 & \underline{ 0.28±0.03447} & 0.46±0.01564 & \textbf{0.49±0.01735} \\
H@19Recall@400 & 0.16±0.02603 & \underline{ 0.20±0.02357} & 0.34±0.01865 & \textbf{0.38±0.01665} \\ \hline
\end{tabular}
}
\label{table:main-tree}
\end{table}

\section{Experiments}
To verify the effectiveness of our method, we conduct experiments to answer the following research questions:
\begin{description}
  \item[RQ1]: How effective is \modelname compared to the state-of-the-art models in the industry? 
  \item[RQ2]: What is the effect of each component of the proposed MISS?
  \item[RQ3]: How does the length of user lifelong sequence affect the performance of \modelname?
  \item[RQ4]: Can our MM-GSU pay more attention to long-term interests? What is the search mechanism of Co-GSU and MM-GSU?
  \item[RQ5]: Can our \modelname drive the growth of online metrics?
\end{description}

\subsection{Experimental Setting}
\subsubsection{Baselines}
We compared with five state-of-the-art models used in industry recommendation systems.
\begin{itemize}
    \item SASRec\cite{kang2018sasrec} is a sequential recommendation model which utilizes self-attention to capture the long-term preferences of users. In our system, SASRec is combined with two-tower paradigm.
    \item TDM\cite{zhu2018tdm} propose tree structures to hierarchically search candidates from coarse to fine and make decisions for each user-node pair using a deep model. In our recommendation system, the deep model of TDM is SIM\cite{pi2020sim}.
    \item TDM+MMoE is the combination of TDM and MMoE.
    \item NANN\cite{chen2022nann} utilizes HNSW\cite{malkov2018hnsw} for candidates searching.
    \item Kuaiformer\cite{liu2024kuaiformer} proposes a transformer-based two tower framework for retrieval. It is a strong baseline in Kuaishou's online recommendation system.  
\end{itemize}

\subsubsection{Metrics}\label{lab:metrics}
To evaluate the performance of retrieval models, we use a widely adopted metric: Recall@K.
For tree-based models, since they would retrieve intermediate nodes during beam search, we propose hierarchical recall to further estimate the effectiveness of the retrieval non-leaf nodes.
Hierarchical recall H@hRecall@K is defined as Recall@K at level $h$ of index tree.
For user $u$, as stated previously, their interacted item set is denoted by $\mathcal{V}_u$ and the retrieval node set at level $h$ during beam search is denoted by $\mathcal{B}_h^{(K)}(u)$.
Suppose $\mathcal{V}_u^{h}=\{\mathbf{Anc}(n;h)|n\in\mathcal{V}_u\}$ to be the set of ancestor nodes at level $h$ of $\mathcal{V}_u$, the hierarchical recall H@hRecall@K is defined as follow:
\begin{align}
    \text{H@hRecall@K}(u,\mathcal{V}_u)= \frac{|\mathcal{V}_u^h\bigcap\mathcal{B}_h^{(K)}(u)|}{|\mathcal{V}_u^h|},
\end{align}

\subsubsection{Dataset}

We tested the model on mass real industrial data. Kuaishou is one of the world's largest short video apps with over \textbf{400 million} daily active users. Kuaishou generates a massive amount of data to support model training and evaluation, with more than \textbf{50 billion} user logs available per day.
To handle such large-scale training data, all models (including the baseline) are optimized using an online learning paradigm. 

To further validate the effectiveness of the model in real industrial scenarios, we will present the results of online A/B tests conducted with real users later in this paper.

\subsubsection{Details} \label{details}
All models, including baselines, are trained in a streaming paradigm. For tree models, we construct a binary tree with 22 levels and apply the following beam search strategy.  
In the main results, Co-GSU has a length of 2000, MM-GSU has 4000 (see details in Sec. \ref{sec:asl}), and ESU has a length of 50.

\subsection{Overall Performance (RQ1)}
\subsubsection{Final retrieval results}
To demonstrate the effectiveness of \modelname in retrieval recommendation, we compare it with five state-of-the-art models, which are practical in the actual industry.
We set sequence length of MM-GSU as 2000 (MISS 2k) and 4000 (MISS 4k).
The experiment is carried out on Kuaishou's real industrial data with recall as the metric. We randomly sample multiply testing datasets and calculate the mean and variance. The result is reported in Table. \ref{table:main}.
From the result, we find that \modelname is state-of-the-art, while "TDM+MMoE" is the second best.
Compared with the second-best results, our method achieves an improvement of $37.93\%$, $30.77\%$ and $47.37\%$ in recall@800, recall@600 and recall@400, respectively.
On average, our method achieves an improvement of $38.69\%$, which evidently confirms the effectiveness of our method.

\subsubsection{Intermediate retrieval results}
To demonstrate effectiveness, we evaluate model performance using hierarchical recall as described in Sec. \ref{lab:metrics}. 
Beam search method in Sec. \ref{details} allows us to analyze hierarchical recall performance, specifically the H@hRecall@K metric at layers 13, 16, and 19. As shown in Table \ref{table:main-tree}, our method significantly outperforms baseline models. Notably, for H@13Recall@800, H@16Recall@800, and H@19Recall@800, our model achieves hierarchical recall increases of $11.84\%$, $52.38\%$, and $59.38\%$, respectively.

\subsection{Ablation Study (RQ2)}\label{abl}
In this section, we examine the impact of our three key modules: MM-GSU, Co-GSU and the multi-modal index tree.
We progressively remove each module from the full model and at each step we can verify the impact of the module.
To be specific, we have the following four versions:
1) full model; 2) version1: full model w/o MM-GSU; 3) version2: version1 w/o Co-GSU, 4) version3: version2 w/o multi-modal index tree. We build an index tree using ID embedding as an alternative.

We evaluate each model variant using hierarchical recall of levels 13, 16, 19 and 22 (leaf).
The results are summarized in Figure. \ref{fig:ablation}.
From the results, it is obvious that removing each module would lead to performance degradation at the four levels, which demonstrates the effectiveness of each module.

\begin{figure}[t]
    \centering
    \includegraphics[width=0.9\linewidth]{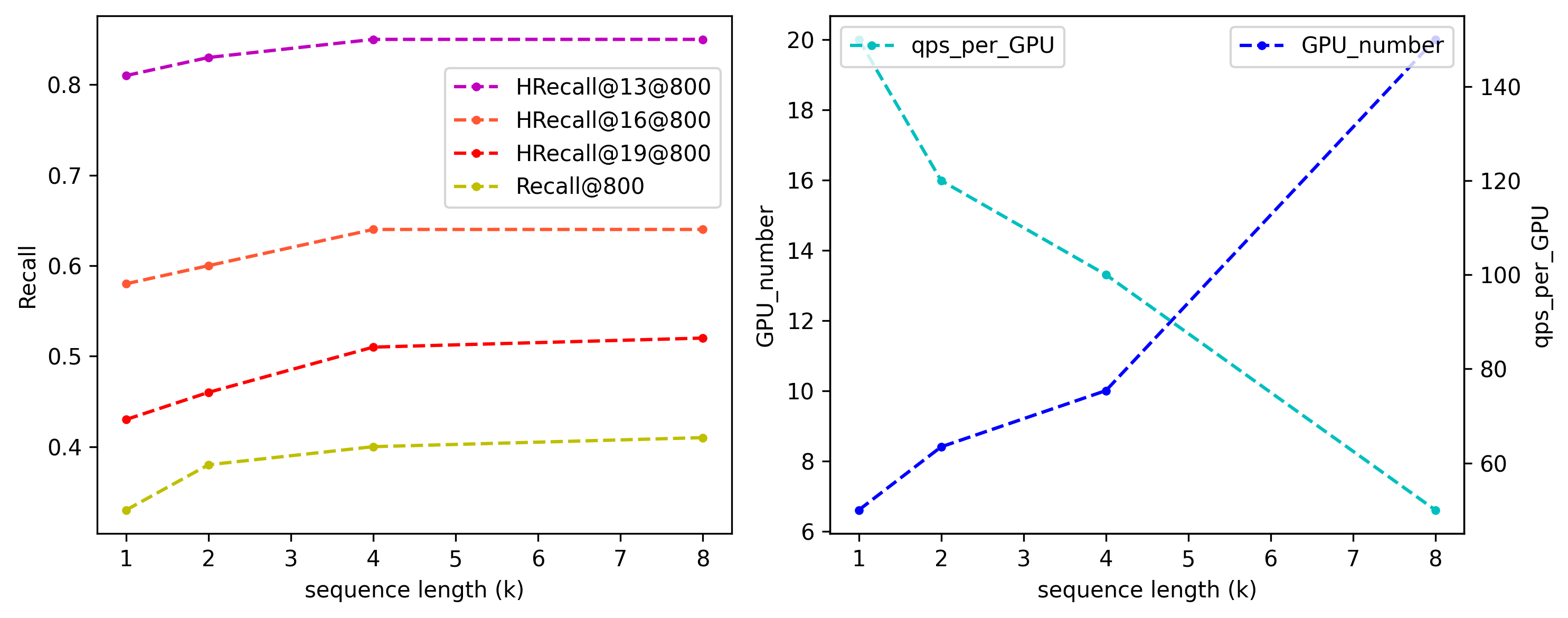}
    \caption{In-depth analysis of the length of user behavior sequence. The left figure shows the relation between recall and sequence length. The right figure shows the relation between computation resource and sequence length.}
    \label{fig:hyperparam}
\end{figure}

\begin{figure}[t]
    \centering
    \includegraphics[width=0.45\textwidth]{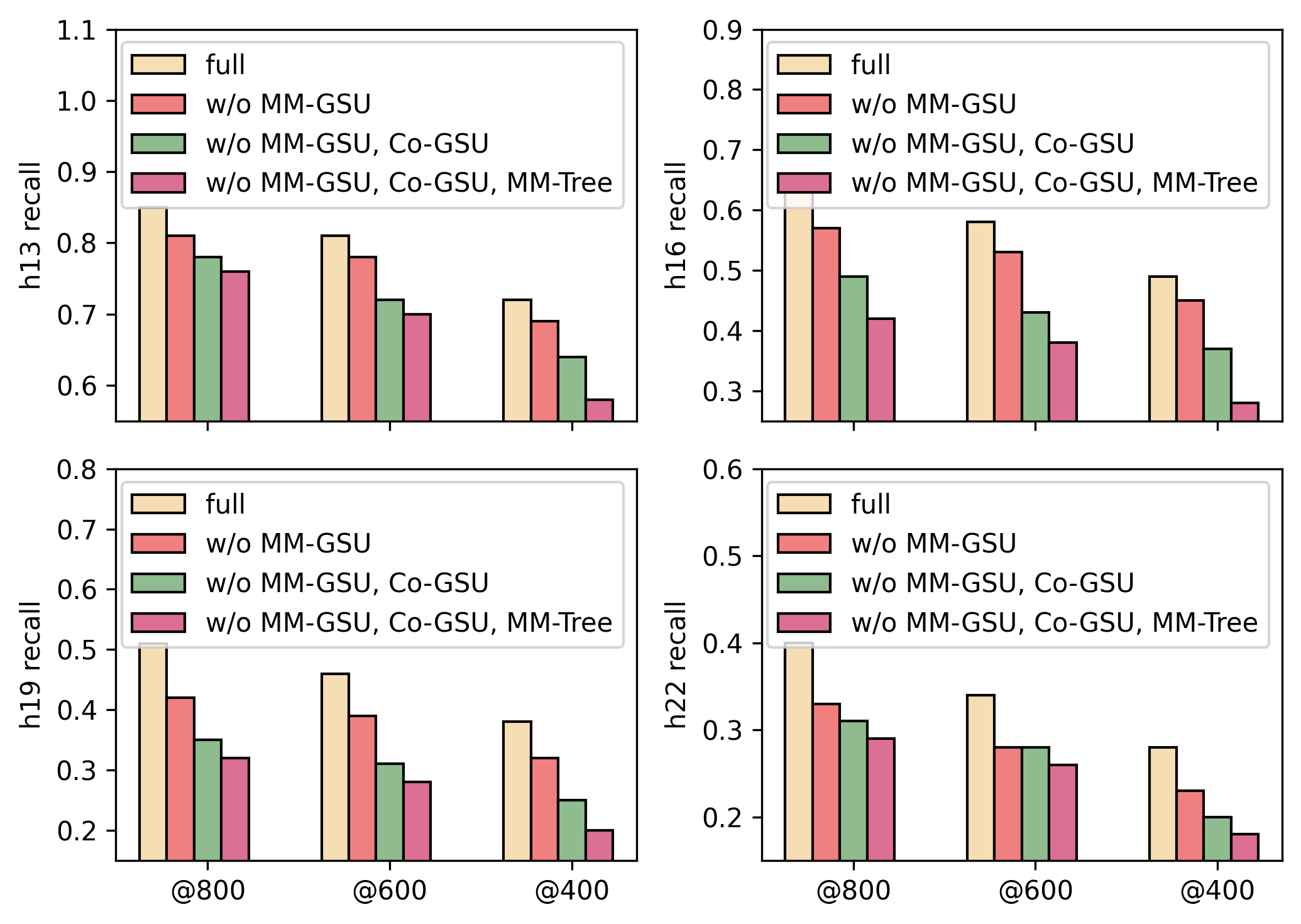}
    \caption{Ablation study of four variants using hierarchical recall metric.}
    \label{fig:ablation} 
\end{figure}

\begin{figure*}[htbp]
    \centering
    \includegraphics[width=\textwidth]{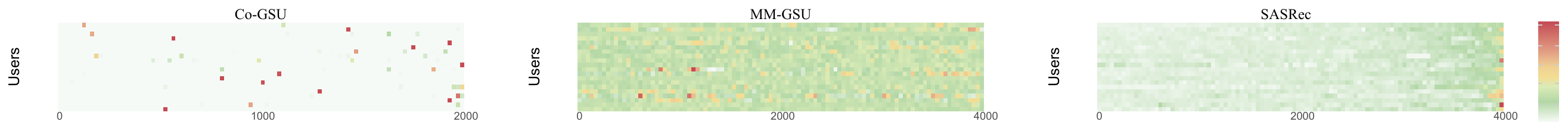}
    \caption{Softmax attention scores during the General Search Unit (GSU) phase for Co-GSU (sequence length of 2000) and MM-GSU (sequence length of 4000), juxtaposed with the self-attention score of SASRec (sequence length of 4000).}
    \label{fig:atten_map} 
\end{figure*}
\begin{figure}[htbp]
    \centering
    \includegraphics[width=0.49\textwidth]{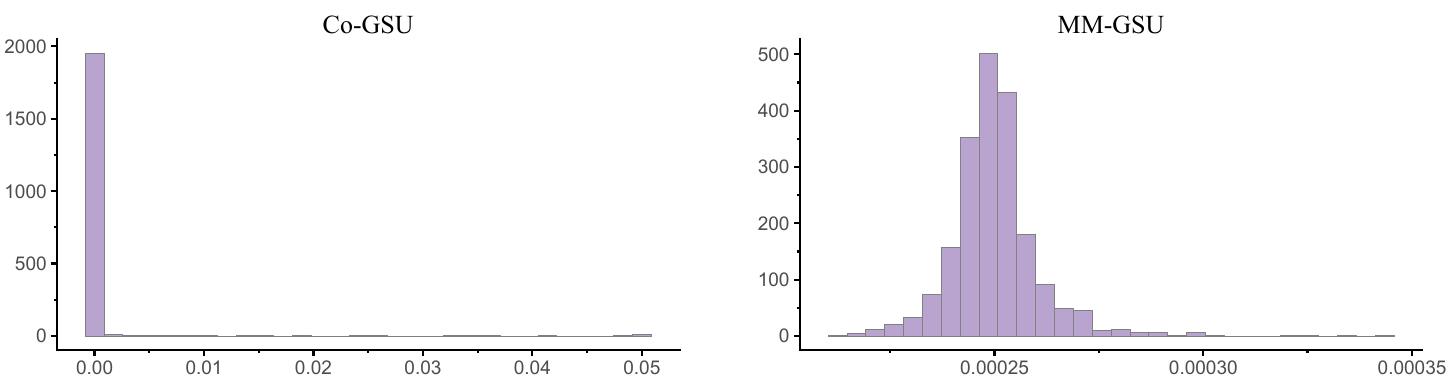}
    \caption{Histograms of Softmax attention scores. Very few items in user sequence can reach high attentions scores in Co-GSU, while in MM-GSU, various locations within the sequence have more equal opportunity to achieve high scores.}
    \label{fig:atten_map_hist} 
\end{figure}

\subsection{In-depth Analysis (RQ3-RQ4)}
\subsubsection{Analysis of Sequence Length} \label{sec:asl}
We analyze the relationship between sequence length, performance, and computational resource requirements.
Specifically, we assess the number of GPUs needed (\textit{GPU\_number}) and the queries per second each GPU can handle (\textit{qps\_per\_GPU}). We tested four sequence lengths of MM-GSU: 1k, 2k, 3k, and 4k, with results summarized in Fig. \ref{fig:hyperparam}.

Notably, there is a positive correlation between sequence length and both performance and GPU resource usage, indicating a trade-off between computational efficiency and model performance. For instance, increasing the sequence length from 4k to 8k yields a slight performance increase (recall@800 from 0.4 to 0.41) but significantly raises resource demand (\textit{GPU\_number} from 10 to 20, \textit{qps\_per\_GPU} from 100 to 50). Thus, we ultimately set the sequence length to 4k to balance efficiency and performance.

\label{sec:analysis}
\subsubsection{Analysis of Attention Scores} 
We visualized the attention scores of Co-GSU (sequence length 2000) and MM-GSU (sequence length 4000) during the general search unit phase of MISS. These scores were compared with the first block self-attention scores from SASRec\cite{kang2018sasrec} (sequence length 4000), as shown in Fig. \ref{fig:atten_map}. Each search method sampled real viewing histories from 20 users, displayed across different rows of the attention map. The leftmost sections represent older viewed items, while the rightmost sections indicate more recent ones.

Notably, items with high attention scores in SASRec are mainly clustered on the right side of the figure, reflecting recent user views, while long-term interests are largely overlooked. In contrast, the GSU, by incorporating query items, enables search results to emerge from various positions within the sequence, rather than being restricted to the most recently viewed items. This feature allows the GSU search mechanism to better utilize the rich information in users' long sequences during the recall phase.

The results suggest that the GSU's search mechanism, particularly in utilizing long-term user sequences, outperforms the dual-tower model's self-attention. This is mainly due to the lack of personalized queries in the dual-tower self-attention, which limits its ability to effectively search for distant tokens.

\subsubsection{Analysis of Two GSU}
When comparing MM-GSU with Co-GSU, it is evident that Co-GSU exhibits significant score disparities within a single sequence (see Fig. \ref{fig:atten_map}), resulting in high attention score positions being relatively sparse. As shown in Fig. \ref{fig:atten_map_hist}, the distribution of softmax attention scores for Co-GSU is highly uneven, with only a small number of tokens achieving high attention scores.

In contrast, MM-GSU displays a denser distribution of high attention score locations (see Fig. \ref{fig:atten_map}), suggesting that various locations within the sequence have more equal opportunity to achieve high scores. This feature enables multi-modal retrieval to provide a more equitable chance for videos within the sequence to be searched, thereby improving the overall accuracy of the model.

More generally, we analyzed the overlap rate of retrieval results between the real samples in Co-GSU and MM-GSU, which was found to be \textbf{only 13\%}. This indicates that the two systems complement each other effectively.

\begin{table}[t]

\centering
\caption{Online A/B Test Result. Confidence intervals (CI) are calculated with 0.05 significance level. }
    \label{tab:ab} 
    \resizebox{\linewidth}{!}{
    \begin{tabular}{l|cc|cc|c}
    \hline
        ~ & \textbf{\makecell{Total App \\ Usage \\ Time} } & \textbf{\makecell{Total App \\ Usage \\ Time (CI)}} & \textbf{\makecell{App Usage \\ Time Per \\ User}} & \textbf{\makecell{App Usage \\\ Time Per \\ User (CI)}} & \textbf{\makecell{Video \\ Watch \\ Time}}\\ \hline
        NANN & +0.103\% & [0.04\%, 0.17\%] &+0.081\% & [0.02\%, 0.14\%] & +0.234\%  \\ 
        Kuaiformer & +0.118\% & [0.03\%, 0.21\%] & +0.080\% & [0.01\%, 0.15\%] & +0.286\%  \\
        TDM {\footnotesize (w/o MM tree)} & +0.121\% & [0.03\% , 0.21\%] & +0.085\% & [0.02\% , 0.15\%] & +0.132\%  \\ 
        TDM {\footnotesize (w/o MM searching)} & +0.166\% & [0.02\%, 0.31\%] & +0.136\% & [0.02\%, 0.25\%] & +0.387\%  \\ \hline
        \textbf{Poposed} & \textbf{+0.248\%} & \textbf{[0.16\%, 0.34\%]} & \textbf{+0.212\%} & \textbf{[0.14\%, 0.28\%]} & \textbf{+0.584\%}  \\ \hline

    \end{tabular}\vspace{-0.3cm} }
\end{table}

\subsection{Online A/B Test Result (RQ5)}

To evaluate the online performance of \modelname, we conduct strict online A/B tests on Kuaishou's video recommendation scenarios. For each recall model, we evaluated its performance within the single-page recommendation system of Kuaishou Lite. Each experiment was conducted over a period of seven days, involving 5\% of the total user. We compared the experimental group against the baseline group in terms of Total App Usage Time and App Usage Time Per User, assessing the growth rates and their corresponding 95\% confidence intervals. Additionally, we analyze the increase in video watch time.

Shown in Tab. \ref{tab:ab}, the experimental results indicate that the recall model incorporating proposed multimodal tree construction and multimodal search modules achieved the best performance. Specifically, we observed a growth of \textbf{0.248\% in Total App Usage Time} and a \textbf{0.212\% increase in App Usage Time Per User}, with the lower bound of the confidence interval also being the highest among the models tested.

\begin{figure}[htbp]
    \centering
    \includegraphics[width=0.50\textwidth]{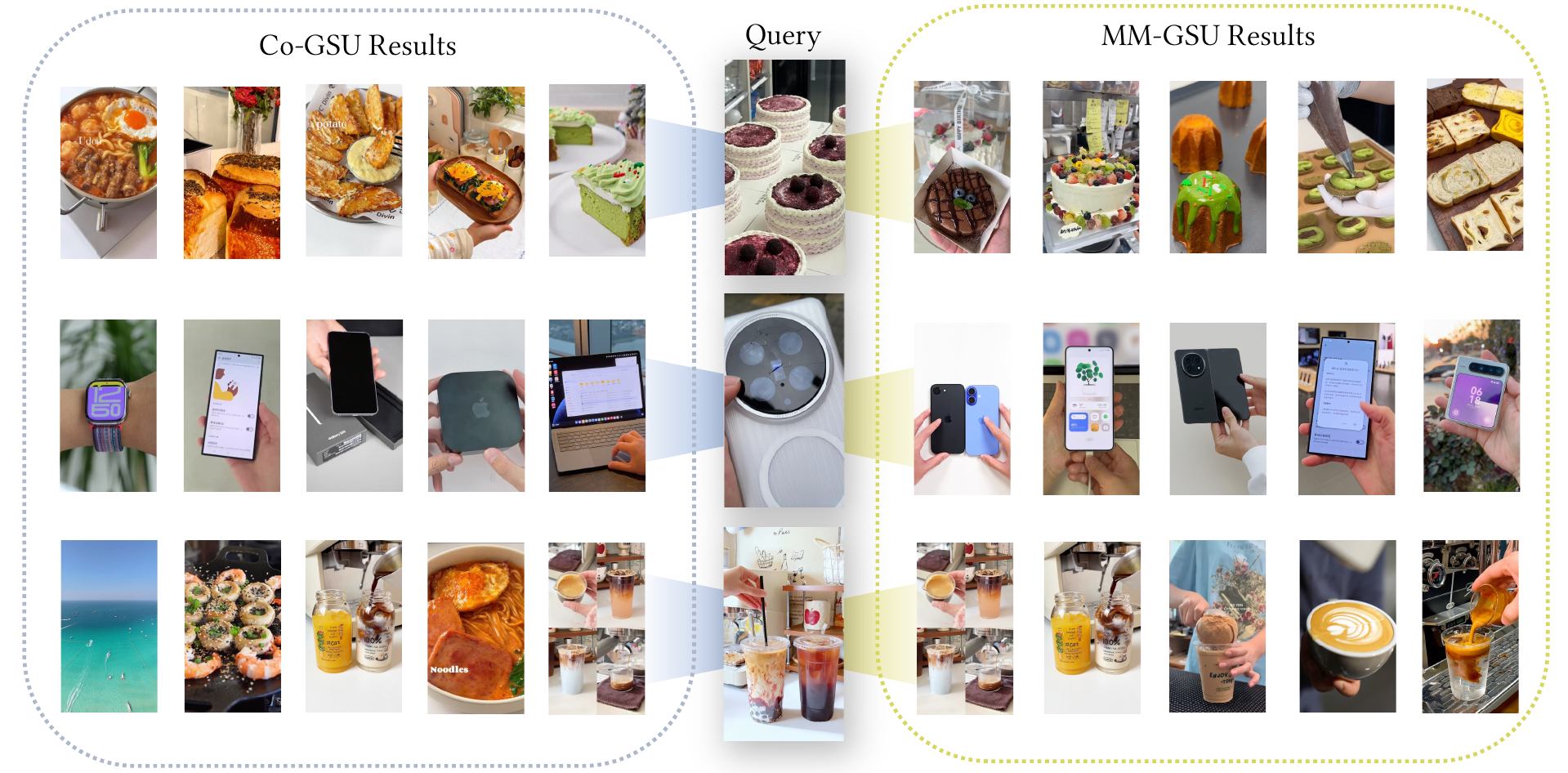}
    \caption{Three specific cases of how searching results can be different between Co-GSU and MM-GSU. In the middle are three query videos. The left column shows searching results of Co-GSU from the user sequence. The right column lists results of MM-GSU.}\vspace{-0.5cm} 
    \label{fig:demo}
\end{figure}

\subsection{Case Study}
This section presents three case studies highlighting the different search results of MM-GSU and Co-GSU. We analyze three query videos shown to users and apply both search methods based on their viewing histories, with Top 5 results displayed in Fig. \ref{fig:demo}.

In the first case, a vlog about a cake baking class yields Co-GSU results that mix general cooking videos, with no specific focus on baking classes. In contrast, MM-GSU's Top 5 results are all centered on cake baking, primarily related to baking classes.

In another case about smartphone fragility, Co-GSU includes various electronics due to co-occurrence learning, which may introduce bias. MM-GSU, however, retrieves semantically relevant smartphone videos.

Lastly, for a coffee-making video, Co-GSU again shows learned co-occurrence, retrieving content beyond coffee, whereas MM-GSU accurately identifies videos specifically about coffee making, demonstrating its contextual relevance.

\section{Conclusion}
In this paper, we represent the pioneering exploration of leveraging multi-modal information within the advanced tree-based retrieval model.
Specifically, for better index structure, we propose multi-modal index tree, which is built using multi-modal embedding to precisely represent video similarity.
To precisely capture diverse user interests in user lifelong sequence, we propose Co-GSU and MM-GSU.
We conducted a wide range of experiments and our method achieves comparable performance.
In the future, we will explore the integration of multi-modal information with other index structure.

\bibliographystyle{ACM-Reference-Format}
\bibliography{sample-base}










\end{document}